\documentclass[prl,aps,twocolumn,showpacs,superscriptaddress,amsmath,amssymb]{revtex4}

\usepackage{graphicx}
\usepackage{psfrag}
\usepackage{epsfig}
\usepackage{color}
\usepackage{subfigure}
\definecolor{dred}{rgb}{0.7,0.0,0.0}
\usepackage{dcolumn}
\usepackage{bm}

\begin{document}

%
%

\title{Nematic State of the Pnictides Stabilized by the \\
Interplay Between Spin, Orbital, and Lattice Degrees of Freedom}

\author{Shuhua Liang} 
\author{Adriana Moreo}
\author{Elbio Dagotto}
 
\affiliation{Department of Physics and Astronomy,University of Tennessee,
Knoxville, TN 37966, USA} 
\affiliation{Materials Science and Technology Division,
Oak Ridge National Laboratory,Oak Ridge, TN 37831, USA}

\date{\today}

\begin{abstract}
{ The nematic state of the iron-based superconductors 
is studied in the undoped limit of the three-orbital ($xz$, $yz$, $xy$) 
spin-fermion model via the introduction of lattice degrees of freedom.  
Monte Carlo simulations show that in order to
stabilize the experimentally observed lattice distortion 
and nematic order, and to reproduce photoemission
experiments, {\it both} the spin-lattice and orbital-lattice couplings are needed.
The interplay between their respective coupling strengths 
regulates the separation between the structural and N\'eel
transition temperatures. Experimental results for the temperature dependence of the 
resistivity anisotropy and the angle-resolved photoemission (ARPES) 
orbital spectral weight are 
reproduced by the present numerical simulations. 
}
\end{abstract}
 
 
\maketitle

{\it Introduction.-} The discovery of high temperature superconductivity 
in the iron-based pnictides and selenides has provided a novel 
playground where several simultaneously active
degrees of freedom (d.o.f.) 
determine the complex properties of these materials~\cite{Fe-SC,pengcheng}. 
The mechanism that leads to superconductivity in these compounds
will only be fully understood once the spin, orbital, lattice, 
and charge are all together considered in a consistent theory.
The parent compounds of most pnictides
become antiferromagnetic (AFM) at a N\'eel temperature $T_N$ where 
long-range collinear spin order develops with wavevector ($\pi$,0) 
in the iron sublattice notation~\cite{pengcheng} breaking rotational symmetry 
from $C_4$ to $C_2$. This state is also 
characterized by an orthorhombic ($\mathcal{O}_{rth}$) lattice distortion with
the longer (shorter) lattice constant along the AFM 
[ferromagnetic (FM)] direction and by the ferro-order of the  
$d_{xz}$ and $d_{yz}$ orbitals that otherwise would be
degenerate~\cite{Fe-SC}. 
In materials such as the undoped 122 compounds, the structural and
magnetic transitions occur at the same temperature. 
However, neutron studies performed on 
LaO$_{1-x}$F$_x$FeAs~\cite{pengcheng} indicate that the AFM transition
can be preceded by a structural transition
at a temperature $T_S>T_N$~\cite{fernandes3,nandi}.

There are two main proposals to explain these results:
(i) In one scenario, the magnetic interactions play the key role
~\cite{spin1,spin2,spin3,fernandes1,fernandes2}. 
In this context the ``nematic''
state~\cite{fradkin} at $T_S$ is induced by breaking the Z$_2$ symmetry 
that links the otherwise degenerate $(\pi,0)$ and $(0,\pi)$ 
collinear states, while at $T_N$ the remaining
$O(3)$ symmetry is broken. However, explicit Monte Carlo (MC) 
calculations using
purely spin models~\cite{MC1,MC2}  revealed only a tiny difference between 
the two critical temperatures. This suggests 
that other d.o.f. may be needed to reinforce
the nematicity mechanism since recent experiments revealed
a nematic transition well above $T_{N}$ for BaFe$_2$As$_2$~\cite{kasahara}
and NaFeAs~\cite{wang-nema} that
persists into the doped regime far from magnetic transitions.
(ii) In another scenario, orbital fluctuations 
are the crucial component~\cite{bnl2,orb1,orb2,orb3,kontani,kontani2,orb4},
similarly as in the manganites where orbital order occurs well
above the magnetic critical temperatures~\cite{CMR}.

Both approaches explain some of the experimental data, 
but in practice it is difficult to disentangle the ``driver'' 
and ``passenger'' roles of the different d.o.f.
The electron acoustic-phonon coupling
responsible for standard tetragonal-orthorhombic structural 
transitions naively appears ruled out as a relevant d.o.f. 
because $\delta$=$[(a_x-a_y)/(a_x+a_y)]\approx 0.003$ 
in the pnictides~\cite{kontani,kontani2,fisher} ($a_x,a_y$=lattice constants), 
and this $\delta$ is considered too small to produce 
the sizable anisotropies experimentally observed~\cite{fisher,kasahara}.

The purpose of this Letter is to revisit
the influence of the lattice d.o.f. in the pnictides via its introduction
into the spin-fermion (SF) model for these materials~\cite{kruger,BNL,shuhua}. 
This model phenomenologically
considers the growing body of experimental evidence that requires a
mixture of itinerant and localized d.o.f. to properly 
address the iron superconductors~\cite{pengcheng,loca1,loca2}. 
Here the itinerant sector will involve
electrons in the $xz$, $yz$, and $xy$ $d$-orbitals~\cite{three}.
The localized spins represent the spin of the other $d$-orbitals~\cite{kruger,BNL} 
or alternatively, in a Landau-Ginzburg context, the magnetic order parameter. 
To our knowledge this is the first time that all these ingredients 
are simultaneously studied, and the complexity of the problem
requires a computational analysis. Moreover, our numerical
approach also allows us to study temperatures above $T_S$ where all d.o.f. 
develop only short-range fluctuations ~\cite{spin3,egami}, a regime difficult to
reach by standard mean-field procedures. Our main result is that a complete
description of the phenomenology of the undoped Fe-based 
superconductors requires
the simultaneous presence of both the spin- and orbital-lattice couplings, 
suggesting a degree of complexity in these materials that was not
previously anticipated.


{\it Model and Method.-} The lattice SF model considered here is based
on the purely electronic model studied before~\cite{kruger,BNL,shuhua} 
supplemented by the coupling to the lattice:
\vspace{-0.2cm}
\begin{equation}
H_{\rm SF} = H_{\rm Hopp} + H_{\rm Hund} + H_{\rm Heis} + H_{\rm SL} + H_{\rm OL} + H_{\rm Stiff}.
\label{ham}
\end{equation}
\vspace{-0.3cm}

This (lengthy) full Hamiltonian is written explicitly in the Supplementary 
Material. 
$H_{\rm Hopp}$ is the Fe-Fe hopping of electrons with the
amplitudes selected in previous publications
to reproduce ARPES results [the specific hopping amplitudes used here 
can be read in Eqs.(1-3) and Table 1 
of Ref.~\cite{three}]. The average number of electrons 
per itinerant orbital is $n$=4/3~\cite{three}. 
Our focus on the undoped case is justified: this limit already contains
the physics under discussion, 
calculations are simpler than for the doped case, 
and the quenched
disordering effect of chemical doping is avoided.
The Hund interaction is canonical:
$H_{\rm Hund}$=$-{J_{\rm H}}\sum_{{\bf i},\alpha} {{{\bf S}_{\bf i}}\cdot{{\bf s}_{{\bf i},\alpha}}}$,
with ${{\bf S}_{\bf i}}$ (${\bf s}_{{\bf i},\alpha}$) 
the localized (itinerant with orbital index $\alpha$) spin. 
$H_{\rm Heis}$ is the Heisenberg interaction 
among the localized spins involving
nearest-neighbors (NN) and next-NN interactions with couplings $J_{\rm NN}$
and $J_{\rm NNN}$, respectively, 
and a ratio $J_{\rm NNN}$/$J_{\rm NN}$=2/3~\cite{shuhua} 
that favors collinear order
(any value larger than 1/2
would have been equally effective). 

Our emphasis will be on the coupling of spin and orbital with the 
structural transition. 
Within the spin-driven scenario, 
the state between $T_N$ and $T_S$ is characterized by short-range 
spin correlations $\Psi_{\bf i}$=$ {\bf{S}_{\bf i}.\bf{S}_{{\bf i}+{\bf x}}} 
- {\bf{S}_{\bf i}.\bf{S}_{{\bf i}+{\bf y}}}$ that satisfy 
$\langle \Psi \rangle$$<$$0$~\cite{fernandes2}, 
where ${\bf{S}_{\bf i}}$ is the spin
of the iron atom at site ${\bf i}$ and ${\bf x,y}$ are unit vectors along the axes.
This spin-nematic phase has been studied analytically both in strong~\cite{spin1,spin2,si} 
and weak coupling~\cite{fernandes1}. 
The $\mathcal{O}_{rth}$-distortion $\epsilon_{\bf i}$
associated to the elastic constant $C_{66}$ 
will be considered here. 
This distortion is produced by coupling
of lattice with the short-range 
magnetic fluctuations via
$H_{\rm SL}$=$-g\sum_{\bf i}\Psi_{\bf i}\epsilon_{\bf i}$~\cite{fernandes1,fernandes2,chu}.
Here, $g$ is the lattice-spin coupling, 
$\epsilon_{\bf i}$ is the $\mathcal{O}_{rth}$ strain
\vspace{-0.2cm}
\begin{equation}
\epsilon_{\bf i}= {1\over{4\sqrt{2}}}\sum_{\nu=1}^4 (|\delta_{\bf i,\nu}^y|-|\delta_{\bf i,\nu}^x|), 
\vspace{-0.1cm}
\label{epsilon}
\end{equation}
and $\delta_{\bf i,\nu}^x$($\delta_{\bf i,\nu}^y$) 
is the component along $x$ ($y$) of the distance 
between the Fe atom at site ${\bf i}$ of the lattice 
and one of its four neighboring  As atoms that are labeled
 by the index $\nu$~\cite{footep}. 
In this context, if the atoms could not move, 
the structural distortion would not occur but the $C_4$ symmetry would 
still spontaneously break at a temperature $T^*>T_N$, leading
to an anisotropic resistivity~\cite{fisher}. 
The spin in $H_{\rm SL}$ will only 
be the localized spin for computational simplicity.
From the other perspective, the orbital fluctuation theory attributes 
the structural transition to the coupling of the lattice
to the $\mathcal{O}_{rth}$ quadrupole operator via
$H_{\rm OL}$=$\lambda\sum_{\bf i}\Phi_{\bf i}\epsilon_{\bf i}$.
Here, $\lambda$ is the orbital-lattice coupling, 
$\Phi_{\bf i}$=$n_{{\bf i},xz}$-$n_{{\bf i},yz}$ is the orbital 
order parameter, and $n_{{\bf i},\alpha}$ the electronic density at site
${\bf i}$ and orbital $\alpha$~\cite{kontani,kontani2}.

\vspace{-0.2cm}
Finally, $H_{\rm Stiff}$ is 
\begin{equation}\begin{split}
H_{\rm Stiff} = {1\over{2}}k\sum_{\bf i}\sum_{\nu=1}^4(|{\bf R}^{\bf i \nu}_{Fe-As}|-R_0)^2+\\
+k'\sum_{<{\bf ij}>}[({a_0\over{R^{\bf ij}_{Fe-Fe}}})^{12}-
2({a_0\over{R^{\bf ij}_{Fe-Fe}}})^6].
\label{Hstiff}
\end{split}\end{equation}
\vspace{-0.3cm}
\noindent The first term 
in Eq.~(\ref{Hstiff}) is the standard harmonic energy. 
The second term contains anharmonic contributions to improve
the simulations' convergence~\cite{foot}. 

Only the $\mathcal{O}_{rth}$-distortion is considered here since our aim is 
to study the structural transition of the parent compounds~\cite{kontani2}.
In equilibrium, the Fe atoms 
form a square lattice with sites labeled by ${\bf i}$ 
and with lattice parameter $a_0$; the As atoms are  
at the center of each plaquette, 
identified with the indices (${\bf i},\nu$),  
with coordinate $z$=$\pm a_0/2$ in alternating plaquettes so that 
the Fe-As equilibrium distance 
is $R_0$=$\sqrt{3}a_0/2$. In our study, each As atom 
is allowed to move in the $x-y$ plane to a new position 
${\bf R}^{\bf i \nu}_{Fe-As}=(\delta_{\bf i,\nu}^x,\delta_{\bf i,\nu}^y,\pm a_0/2)$ 
with respect to the Fe atom that was at site ${\bf i}$ when in equilibrium. 
The distance between Fe atoms, $R^{\bf ij}_{Fe-Fe}$, 
is determined {\it globally} via the variables $a_x$ and $a_y$, 
both equal to $a_0$ when in equilibrium, satisfying the 
constraints 
$2Na_x=\sum_{{\bf i}=1}^N\sum_{\nu} |\delta_{{\bf i},\nu}^x|$ and $2Na_y=\sum_{{\bf i}=1}^N \sum_{\nu}|\delta_{{\bf i},\nu}^y|$ where 
$N$ is the number of sites and $\nu$=1,...,4 are the 
four As atoms connected to each Fe. Note that this procedure is qualitatively different
from studies of Jahn-Teller distortions in Mn-oxides where the Mn-Mn
distance was fixed~\cite{CMR}, while here the Fe-Fe distances can change 
due to the  $\mathcal{O}_{rth}$-distortion leading to the global adjustments in lattice spacings.

The Hamiltonian is here studied via a standard MC 
simulation in the classical (a) localized spins 
${\bf S_i}$ and (b) atomic
displacements  $\delta_{\bf i,\nu}^x$ and $\delta_{\bf i,\nu}^y$. 
For each MC configuration of spins and atomic positions 
the fermionic quantum Hamiltonian is diagonalized via library subroutines,
as extensively discussed in the manganite context~\cite{CMR}, rendering the
study computationally demanding. 

{\it Results.-} The MC simulations were performed on $8 \times 8$ 
square clusters using ``twisted boundary conditions'' that effectively 
reduce finite size effects, as discussed before~\cite{shuhua}.
Typically 8,000 MC steps were used for thermalization 
and 50,000-100,000 steps for measurements at each temperature $T$ and 
for each set of parameters. The Hund interaction was set to $J_{\rm H}$=$0.1$ eV, 
and the classical Heisenberg couplings to $J_{\rm NN}$=$0.012$ eV
and $J_{\rm NNN}$=$0.008$ eV, similarly as in Ref.~\cite{shuhua}. Fixing
some parameters to values used in previous investigations simplifies the analysis and 
allow us to focus on the effects of the lattice into a previously studied system. 
The stiffness constants were selected so that the dimensionless 
couplings $\tilde\lambda$=${2\lambda\over{kW}}$ 
and $\tilde g$=${2g\over{kW}}$~\cite{CMR} are experimentally 
realistic~\cite{f3} ($W$=fermionic bandwidth). Calculations indicate that both parameters 
should be smaller than 1 in pnictides~\cite{kontani,boeri,spin3,kontani2}.
The magnetic transition will be determined by the magnetic susceptibility 
\vspace{-0.2cm}
\begin{equation}
\chi_{S(\pi,0)}=N\beta\langle S(\pi,0)-\langle S(\pi,0)\rangle\rangle^2,
\label{Xs}
\end{equation}
\vspace{-0.1cm}
\noindent where $\beta=1/k_BT$, $N$ is the number of lattice sites, 
and $S(\pi,0)$ is the magnetic structure factor [at
the wavevector $(\pi,0)$ of relevance in pnictides] 
obtained via the Fourier transform of the real-space 
spin-spin correlations measured during the simulations. The 
structural transition is determined by the behavior of the 
lattice susceptibility defined by
$\chi_{\delta}$=$N\beta\langle \delta-\langle\delta\rangle\rangle^2$, 
where $\delta={(a_x-a_y)\over{(a_x+a_y)}}$.

\begin{figure}[thbp]
\subfigure {\includegraphics[trim = 8mm 5mm 4mm 5mm,width=0.22\textwidth,angle=0]{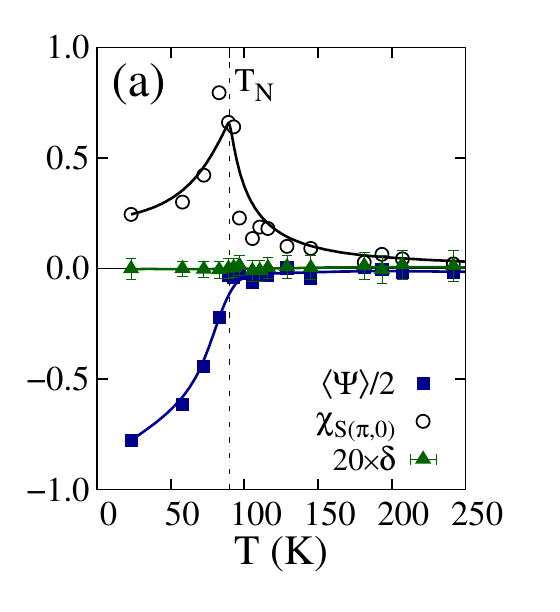}\label{g0l15a}}
\subfigure {\includegraphics[trim = 8mm 5mm 4mm 5mm,width=0.22\textwidth,angle=0]{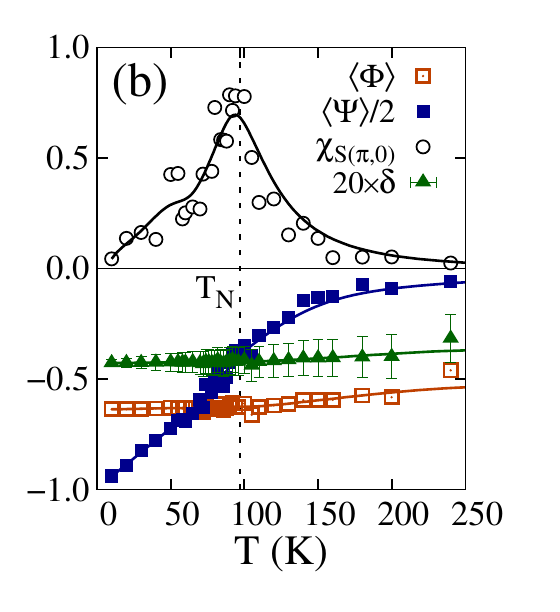}\label{g0l150b}}\\
\caption{(Color online) Spin magnetic susceptibility $\chi_{S(\pi,0)}$ 
(circles), spin-nematic order parameter $\langle \Psi \rangle$ (filled squares), and 
lattice distortion $\delta$ (triangles) vs. $T$ at
$\tilde g=0$ and 
(a) $\tilde\lambda=0.12$ and (b) $\tilde\lambda=1.2$ (in the latter, 
open squares indicate orbital order). 
$T_N$ is indicated by the dashed line.}
\label{g0l15}
\end{figure}

{\it Individual couplings.-} To isolate the individual roles that the spin and 
orbital d.o.f. play in their interaction with the lattice, first
the case $\tilde g=0$ was studied, varying $T$
at several values of $~\tilde\lambda$. At $\tilde\lambda=0.12$ 
neither a sizable lattice distortion [as indicated by the triangles 
in Fig.~\ref{g0l15}(a)] nor orbital order were observed,   
and only a N\'eel transition at $T_N=90$~K into a collinear 
AFM $(\pi,0)$ state was found (see 
circles in the figure). To develop  a more robust
lattice distortion $\tilde\lambda$ must be increased
to unphysical large values.
In fact, numerically it was observed that varying $\tilde\lambda$ the
orbital order and structural distortion are stabilized 
for $\tilde\lambda>0.8$. However, in this $\tilde \lambda$ 
regime, already larger than estimations~\cite{kontani,boeri}, 
the $\mathcal{O}_{rth}$-distortion has the longest lattice constant along
the FM direction (see Fig.~\ref{g0l15}(b) at $\tilde\lambda$=$1.2$), 
qualitatively  opposite to experimental observations~\cite{lambda}.
As a consequence, in our model, that relies on a particular set of
hopping amplitudes chosen to fit ARPES experiments,  the physical 
$\mathcal{O}_{rth}$/magnetic state of pnictides cannot arise from 
short-range orbital fluctuations alone~\cite{kontani2}. 
Let us study next the role played by the spin-lattice coupling 
by setting instead $\tilde\lambda$=$0$ 
and focusing on, e.g., $\tilde g$=$0.16$. In this case, 
 a peak in $\chi_{\delta}$ 
[see Fig.~\ref{g2l0}(a)] denotes a structural transition. This transition now 
has the experimentally correct $\mathcal{O}_{rth}$-distortion below $T_N$,  
i.e. $\delta >0$, and it
occurs simultaneously with the N\'eel transition at $T_S$=$T_N$=$153$~K. 
The ordered phase now has both long-range magnetic order and a 
long-range $\mathcal{O}_{rth}$-distortion with 
$\delta$=$(a_x-a_y)/(a_x+a_y)\approx 0.0037$ (green triangles), remarkably
close to experiments suggesting that
the small couplings to the lattice considered here are physically reasonable.
However, setting $\tilde\lambda=0$ no orbital order 
was observed, at least with the hopping amplitudes employed here. 
Moreover our study shows that $T_N$ remains
equal (within the accuracy of our effort) to $T_S$ in the physical regime, 
contrary to experiments. Then, neither
the limits $\tilde\lambda$=$0$ nor $~\tilde g$=$0$ are sufficient
to fully accommodate the phenomenology of the pnictides.

\begin{figure}[thbp]
\begin{center}
\subfigure {\includegraphics[trim = 8mm 5mm 4mm 5mm,width=0.22\textwidth,angle=0]{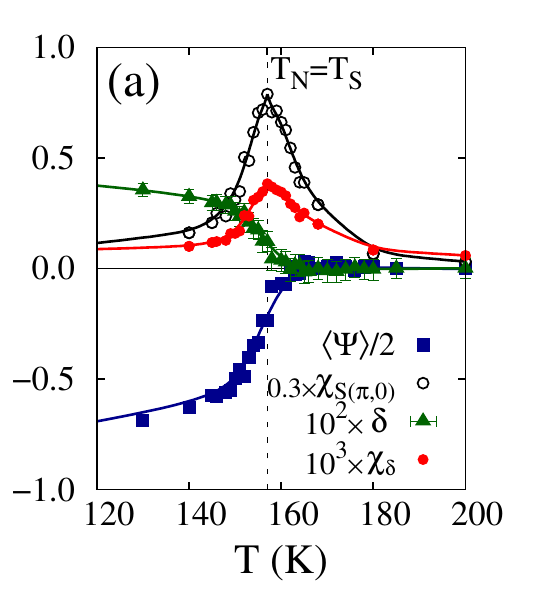}\label{g2l0}}
\subfigure {\includegraphics[trim = 4mm 5mm 8mm 5mm,width=0.22\textwidth,angle=0]{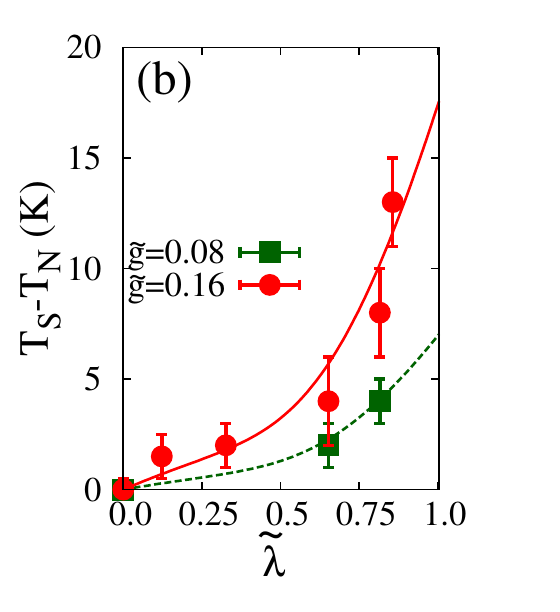}\label{ratio}}\\
\vspace{-0.3cm}
\caption{(color online) (a)  Spin magnetic susceptibility $\chi_{S(\pi,0)}$
(open circles), lattice distortion susceptibility $\chi_{\delta}$ (filled circles), 
spin-nematic order parameter $\langle \Psi \rangle$ (squares), 
and lattice distortion $\delta$ (triangles) 
vs. $T$ for couplings ${\tilde g}$=$0.16$ and
$\tilde\lambda$=$0$. 
$T_N$ and the structural transition temperature $T_S$ are indicated 
by the dashed line. (b) The temperature difference 
between $T_S$ and $T_N$ vs. $\tilde\lambda$, at $\tilde g=0.08$ and 0.16.}
\label{g2l0}
\vspace{-0.3cm}
\end{center}
\end{figure}

\begin{figure}[thbp]
\begin{center}
\vspace{-0.3cm}
\includegraphics[width=8cm,clip,angle=-90]{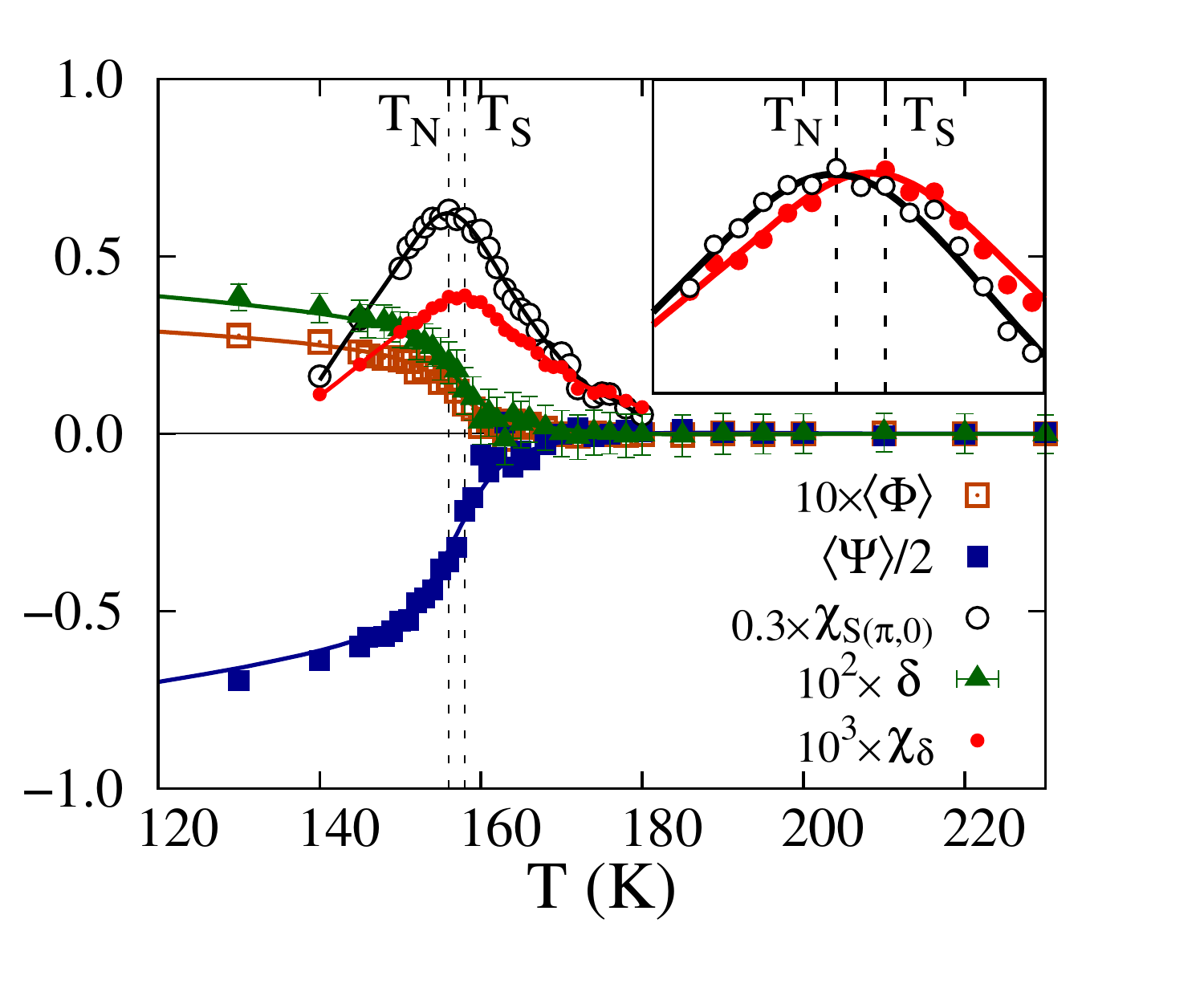}
\vspace{-0.6cm}
\caption{(color online) Spin magnetic susceptibility $\chi_{S(\pi,0)}$  (open 
circles), lattice distortion susceptibility $\chi_{\delta}$ (filled circles), 
spin-nematic order parameter $\langle \Psi \rangle$ (filled squares), orbital 
order $\langle \Phi \rangle$ (open squares), 
and lattice distortion $\delta$ (triangles) vs. $T$ 
at couplings $\tilde g$=$0.16$ and
$\tilde\lambda$=$0.12$. $T_N$ and $T_S$ 
are indicated by the dashed lines. {\it Inset:} close-up of 
the $\chi_{\rm S(\pi,0)}$ and $\chi_{\delta}$ peaks, shifted vertically
for better comparison.}
\label{g2l15}
\vspace{-0.7cm}
\end{center}
\end{figure}

{\it Combined couplings.-} Our main result is that the {\it combined} effect
of the coupling of the lattice to both spins and orbitals is needed 
to reach a regime with all the characteristics of the
states found experimentally in pnictides.
By turning on both the spin- and orbital-lattice interactions our MC studies show
that the structural transition moves to a temperature 
higher than the magnetic transition so that $T_S>T_N$, as
shown in Fig.~\ref{g2l0}(b) at $\tilde g=0.16$ and $0.08$. 
For small couplings in the experimental range, such 
as  $\tilde\lambda=0.12$ and $\tilde g=0.16$, 
the difference $T_S-T_N$ is concomitantly 
small but it is numerically clear, with $\chi_{\delta}$ systematically
above (below) $\chi_{S(\pi,0)}$ at temperatures above (below) the critical region.
More specifically, $T_N=156$~K from the peak in $\chi_S$ 
(open black circles) in Fig.~\ref{g2l15}, and
$T_S=158$~K from the peak in $\chi_{\delta}$ (filled circles). 
The difference in the position of the two maxima (see inset) 
has been extensively analyzed  
repeating MC runs with different starting configurations and
statistics, and it appears robust. Moreover, $T_S-T_N$ can be further enhanced 
by increasing $\tilde\lambda$ [see Figs.~2(b) and ~9 
(Suppl. Mat.)]~\cite{kontaninew}. 
The intermediate phase has a 
broken $Z_2$ symmetry with short-range 
NN spin-spin correlations characterized by 
$\langle \Psi \rangle$$<$$0$ indicating spin-nematic order (filled squares), 
$\delta>0$ indicating $\mathcal{O}_{rth}$ distortion (triangles), and
$\langle \Phi \rangle$$>$$0$ indicating orbital order (open squares).

\begin{figure}[thbp]
\subfigure {\includegraphics[trim = 8mm 5mm 4mm
5mm,width=0.22\textwidth,angle=0]{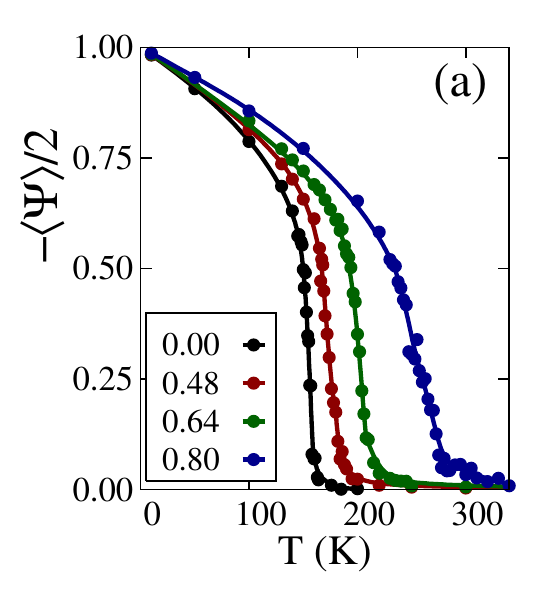}\label{AA}}
\subfigure {\includegraphics[trim = 4mm 5mm 8mm
5mm,width=0.22\textwidth,angle=0]{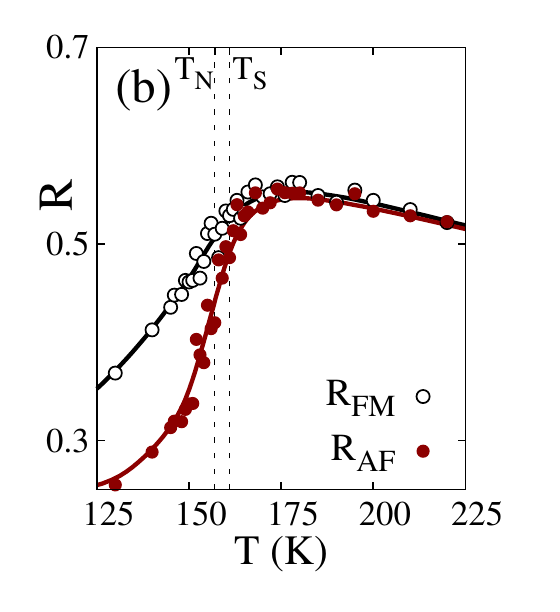}\label{con}}\\
\vspace{-0.1cm}
\caption{(Color online) (a) Spin-nematic order parameter $\langle\Psi\rangle$ 
vs. $T$ at $\tilde g=0.16$ and for the values of $\tilde\lambda$ 
indicated. (b) MC resistance along 
the $x$ (AFM) and $y$ (FM) direction
varying $T$.
Dashed lines indicate $T_N$ and $T_S$ 
at $\tilde g=0.16$ and $\tilde\lambda=0.12$. 
}
\vspace{-0.2cm}
\label{fig4}
\end{figure}

The order of the transitions was also investigated.
In Fig.~\ref{fig4}(a) the spin-nematic order parameter
$\langle\Psi\rangle$ is shown varying $T$ at several
 $\tilde\lambda$'s and fixed $\tilde g$=0.16.
At small $\tilde \lambda$, where $T_N$=$T_S$ according
to Fig.~2(a), the transition is abrupt as in 
a first-order transition. Upon increasing $\tilde \lambda$,
leading to $T_S$$>$$T_N$, the transition becomes 
continuous as in a second-order transition.
This is in agreement with predictions of an
effective low-energy model~\cite{fernandes1}.

{\it Comparison with experiments.-} As in the previous
effort employing the purely electronic SF 
model~\cite{shuhua} the resistance R 
along the AFM and FM directions was calculated varying $T$.
While the reproduction of the 
uniaxial-pressure experimental results~\cite{fisher}  
required previously an explicit anisotropy in the Heisenberg 
couplings to mimick strain, now 
the asymmetry develops {\it spontaneously} as shown 
in Fig.~\ref{fig4}(b). R 
along the FM direction becomes larger than along the AFM direction at $T\approx T_S$
suggesting that the anisotropy observed above $T_S$ in experiments may be due to the 
external strain~\cite{strain1,strain2}. 
 
\begin{figure}
\begin{center}

\makebox[0pt][l]{\includegraphics[trim = 0mm 0mm 0mm 0mm,clip,width=0.48\textwidth,angle=0]{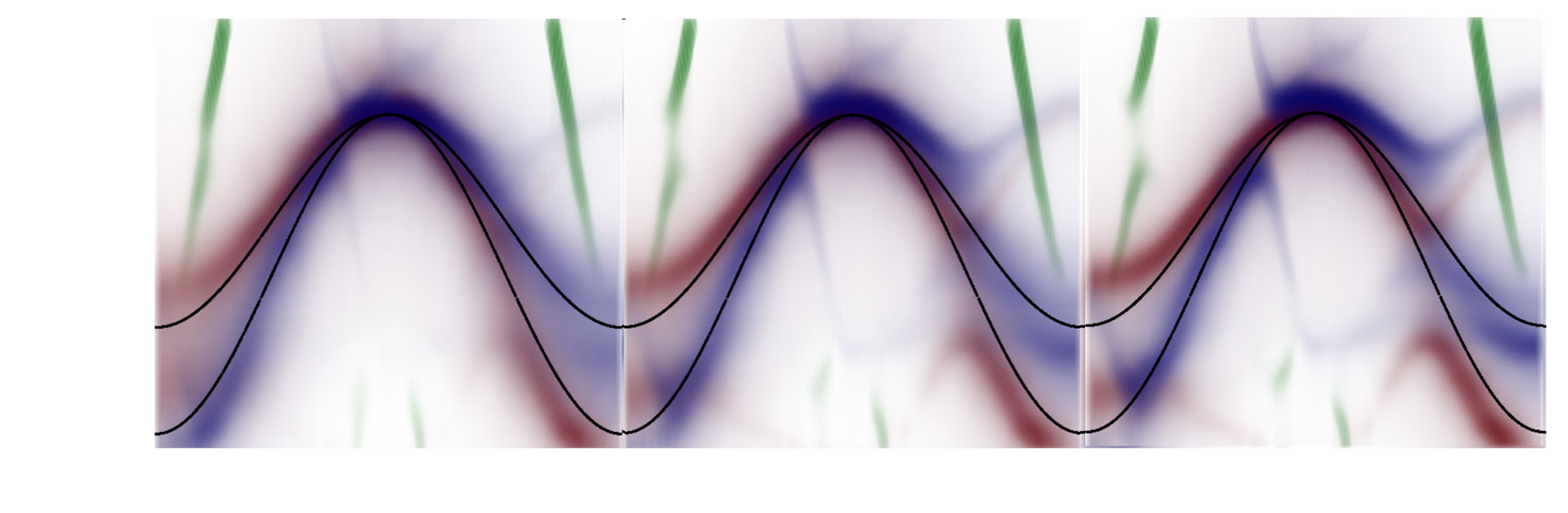}}%
\includegraphics[trim = 0mm 0mm 0mm 0mm,clip,width=0.48\textwidth,angle=0]{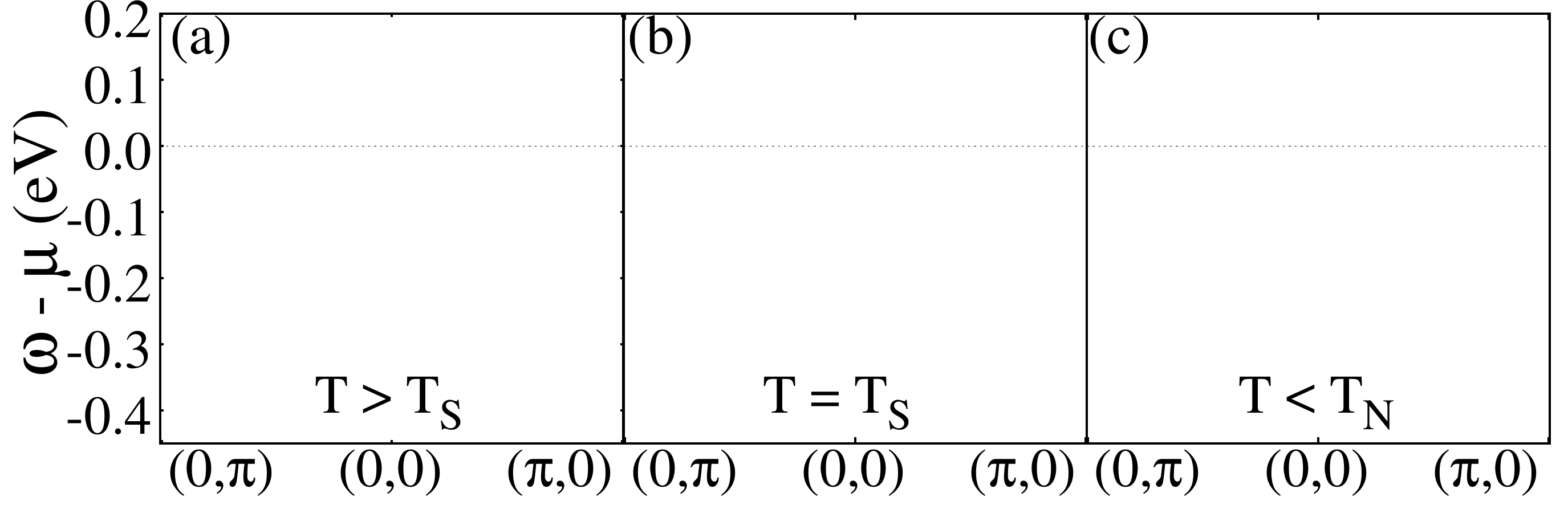}

\vspace{-0.2cm}
\caption{(Color online) Orbital-resolved spectral weight of the SF model
along the directions $(0,\pi)-(0,0)-(\pi,0)$ in momentum
space for (a) $T$=$165$~K, (b) 
$T$=$158$~K, and (c) $T$=$145$~K, at $\tilde g$=$0.16$ and $\tilde\lambda$=$0.12$. 
The non-interacting band dispersion is indicated 
by the solid black lines. The spectral weight for the $d_{xz}$, $d_{yz}$, and
$d_{xy}$ orbitals is indicated by the red, blue, and green dots, respectively.
\label{spectral}}
\vspace{-0.4cm}
\end{center}
\end{figure}

Our study also reproduces
the ARPES experiments~\cite{arpes1,arpes2,arpes3} 
where an asymmetry develops between the spectral weight for 
the $xz$ and $yz$ orbitals along the $\Gamma-X$ and the $\Gamma-Y$ directions upon cooling.
In Fig.~\ref{spectral} it is shown that 
along the $\Gamma-X$ [$\Gamma-Y$] direction, mainly near $(\pi,0)$ [$(0,\pi)$],
the spectral weight 
for the $yz$ ($xz$)
orbital moves closer to (further from) 
the Fermi level as $T$ is
lowered, compatible with the development of 
orbital order with $\langle \Phi \rangle$$>$$0$. 
The asymmetry is obtained here without 
explicit symmetry breaking at the Hamiltonian level~\cite{aniso}.
Note also that orbital order may only occur near the Fermi Surface~\cite{daghofer-OO}. 
It is important to remark
that in spite of the small values of $\tilde \lambda$ and $\tilde g$
used in our effort, their influence is sufficient to create observable consequences
such as the anisotropies in transport and ARPES. In addition, a recent
pair-distribution function analysis reported the presence of robust {\it local} 
$\mathcal{O}_{rth}$-distortions~\cite{egami2}, 
hinting that the lattice d.o.f. 
is  more important than previously believed~\cite{lattice-arpes}.

{\it Conclusions.-} In the model analyzed here,
the couplings of the spin and orbital d.o.f. with the lattice 
are {\it both} important to stabilize  
the state that breaks the $C_4$ 
symmetry above the N\'eel transition. 
The spin-lattice coupling induces 
the correct experimentally observed $\mathcal{O}_{rth}$-distortion, 
while the orbital-lattice coupling 
generates the ARPES-observed orbital order 
and the higher temperature structural
transition. As a consequence, 
our study suggests that the complex nematic properties of
the pnictides parent compounds arise from a
subtle cooperation among all the participating degrees of freedom. 

{\it Acknowledgment.-} This work was supported by the U.S. Department of Energy, 
Office of Basic Energy Sciences, Materials
Sciences and Engineering Division.

\vfill\eject

\vspace{2.0cm}

\section{Supplementary Material}

\subsection{Full Hamiltonian}

The full Hamiltonian of the 
spin-fermion model with lattice interactions incorporated is given by:
\vspace{-0.2cm}
\begin{equation}
H_{\rm SF} = H_{\rm Hopp} + H_{\rm Hund} + H_{\rm Heis} + H_{\rm SL} + H_{\rm OL} + H_{\rm Stiff}.
\label{hamap}
\end{equation}
\vspace{-0.2cm}
\noindent The hopping component is made of three contributions,
\begin{equation}
H_{\rm Hopp}=H_{xz,yz}+H_{xy}+H_{xz,yz;xy}. 
\label{hop}
\end{equation}
\vspace{-0.2cm}
\noindent The first  term involves the $xz$ and $yz$ orbitals as follows:
\begin{equation}\begin{split}
H_{xz,yz}&=
\{-t_1\sum_{{\bf i},\sigma}(d^{\dagger}_{{\bf i},xz,\sigma}
d^{\phantom{\dagger}}_{{\bf i}+\hat y,xz,\sigma}+d^{\dagger}_{{\bf i},yz,\sigma}
d^{\phantom{\dagger}}_{{\bf i}+\hat x,yz,\sigma}) \\
&-t_2\sum_{{\bf i},\sigma}(d^{\dagger}_{{\bf i},xz,\sigma}
d^{\phantom{\dagger}}_{{\bf i}+\hat x,xz,\sigma}+d^{\dagger}_{{\bf i},yz,\sigma}
d^{\phantom{\dagger}}_{{\bf i}+\hat y,yz,\sigma}) \\
&-t_3\sum_{{\bf i},\hat\mu\not=\hat\nu,\sigma}(d^{\dagger}_{{\bf i},xz,\sigma}
d^{\phantom{\dagger}}_{{\bf i}+\hat\mu+\hat\nu,xz,\sigma}+d^{\dagger}_{{\bf i},yz,\sigma}
d^{\phantom{\dagger}}_{{\bf i}+\hat\mu+\hat\nu,yz,\sigma}) \\
&+t_4\sum_{{\bf i},\sigma}(d^{\dagger}_{{\bf i},xz,\sigma}
d^{\phantom{\dagger}}_{{\bf i}+\hat x+\hat y,yz,\sigma}+d^{\dagger}_{{\bf i},yz,\sigma}
d^{\phantom{\dagger}}_{{\bf i}+\hat x+\hat y,xz,\sigma}) \\
&-t_4\sum_{{\bf i},\sigma}(d^{\dagger}_{{\bf i},xz,\sigma}
d^{\phantom{\dagger}}_{{\bf i}+\hat x-\hat y,yz,\sigma}+d^{\dagger}_{{\bf i},yz,\sigma}
d^{\phantom{\dagger}}_{{\bf i}+\hat x-\hat y,xz,\sigma})\\
&+h.c.\}-\mu\sum_{\bf i}(n_{{\bf i},xz}+n_{{\bf i},yz}).
\label{ham12}
\end{split}\end{equation}
\noindent The second term contains the hoppings related with the $xy$ orbital:
\begin{equation}\begin{split}
H_{xy}=&\ t_5\sum_{{\bf i},{\hat \mu},\sigma}(d^{\dagger}_{{\bf i},xy,\sigma}
d^{\phantom{\dagger}}_{{\bf i}+\hat \mu,xy,\sigma}+h.c.)\\
&-t_6\sum_{{\bf i},\hat\mu\not=\hat\nu,\sigma}(d^{\dagger}_{{\bf i},xy,\sigma}
d^{\phantom{\dagger}}_{{\bf i}+\hat\mu+\hat\nu,xy,\sigma}+h.c.)\\
&+\Delta_{xy}\sum_{\bf i}n_{{\bf i},xy}-\mu\sum_{\bf i}n_{{\bf i},xy},\;
\label{ham3}
\end{split}\end{equation}
\noindent Finally, the last term contributing to the hopping is:
\begin{equation}\begin{split}
H_{\rm xz,yz;xy}=&-t_7\sum_{{\bf i},\sigma}[(-1)^{|{\bf i}|}d^{\dagger}_{{\bf i},xz,\sigma}
d^{\phantom{\dagger}}_{{\bf i}+\hat x,xy,\sigma}+h.c.]\\
&-t_7\sum_{{\bf i},\sigma}[(-1)^{|{\bf i}|}d^{\dagger}_{{\bf i},xy,\sigma}
d^{\phantom{\dagger}}_{{\bf i}+\hat x,xz,\sigma}+h.c.]\\
&-t_7\sum_{{\bf i},\sigma}[(-1)^{|{\bf i}|}d^{\dagger}_{{\bf i},yz,\sigma}
d^{\phantom{\dagger}}_{{\bf i}+\hat y,xy,\sigma}+h.c.] \\
&-t_7\sum_{{\bf i},\sigma}[(-1)^{|{\bf i}|}d^{\dagger}_{{\bf i},xy,\sigma}
d^{\phantom{\dagger}}_{{\bf i}+\hat y,yz,\sigma}+h.c.] \\
&-t_8\sum_{{\bf i},\sigma}[(-1)^{|{\bf i}|}d^{\dagger}_{{\bf i},xz,\sigma}
d^{\phantom{\dagger}}_{{\bf i}+\hat x+\hat y,xy,\sigma}+h.c.] \\
&+t_8\sum_{{\bf i},\sigma}[(-1)^{|{\bf i}|}d^{\dagger}_{{\bf i},xy,\sigma}
d^{\phantom{\dagger}}_{{\bf i}+\hat x+\hat y,xz,\sigma}+h.c.] \\
&-t_8\sum_{{\bf i},\sigma}[(-1)^{|{\bf i}|}d^{\dagger}_{{\bf i},xz,\sigma}
d^{\phantom{\dagger}}_{{\bf i}+\hat x-\hat y,xy,\sigma}+h.c.]\\
&+t_8\sum_{{\bf i},\sigma}[(-1)^{|{\bf i}|}d^{\dagger}_{{\bf i},xy,\sigma}
d^{\phantom{\dagger}}_{{\bf i}+\hat x-\hat y,xz,y\sigma}+h.c.] \\
&-t_8\sum_{{\bf i},\sigma}[(-1)^{|{\bf i}|}d^{\dagger}_{{\bf i},yz,\sigma}
d^{\phantom{\dagger}}_{{\bf i}+\hat x+\hat y,xy,\sigma}+h.c.] \\
&+t_8\sum_{{\bf i},\sigma}[(-1)^{|{\bf i}|}d^{\dagger}_{{\bf i},xy,\sigma}
d^{\phantom{\dagger}}_{{\bf i}+\hat x+\hat y,yz,\sigma}+h.c.] \\
&+t_8\sum_{{\bf i},\sigma}[(-1)^{|{\bf i}|}d^{\dagger}_{{\bf i},yz,\sigma}
d^{\phantom{\dagger}}_{{\bf i}+\hat x-\hat y,xy,\sigma}+h.c.] \\
&-t_8\sum_{{\bf i},\sigma}[(-1)^{|{\bf i}|}d^{\dagger}_{{\bf i},xy,\sigma}
d^{\phantom{\dagger}}_{{\bf i}+\hat x-\hat y,yz,\sigma}+h.c.].
\label{ham12_3}
\end{split}\end{equation}
In the equations shown above, the operator $d^{\dagger}_{{\bf i},\alpha,\sigma}$ creates an electron at 
site ${\bf i}$ of the two-dimensional lattice of irons. The orbital index is $\alpha=$  
$xz$, $yz$, or $xy$, and the $z$-axis spin projection is denoted by $\sigma$. The chemical potential
used to regulate the electronic density is $\mu$. The symbols ${\hat x}$ and ${\hat y}$ denote
vectors along the axes that join NN atoms.
The values of the hoppings $t_i$ were discussed originally in Ref.~\cite{three} and for the benefit of
the readers they are reproduced here in Table~\ref{tableap}, including also the value of the energy splitting $\Delta_{xy}$.
\begin{table}
\caption{Values of the parameters that appear in the tight-binding portion of the three-orbital model
  Eqs.(\ref{ham12}) to (\ref{ham12_3}). The overall energy unit is
  electron volts.\label{tableap}}
 \begin{tabular}{|ccccccccc|}\hline
$t_1$ & $t_2$ & $t_3$ & $t_4$ & $t_5$ & $t_6$ & $t_7$ & $t_8$ &
   $\Delta_{xy}$\\
\hline
  0.02   &0.06    &0.03   &$-0.01$&$0.2$ & 0.3 & $-0.2$ & $0.1$&
  0.4\\ \hline
 \end{tabular}
\end{table}

The remaining terms of the Hamiltonian have been presented in the main text, but they are reproduced below again 
for completeness. The symbols $\langle \rangle$ denote NN while $\langle\langle \rangle\rangle$ denote NNN. 
The rest of the notation was already explained in the main text.  
\begin{equation}
H_{\rm Hund}= -{J_{\rm H}}\sum_{{\bf i},\alpha} {{{\bf S}_{\bf i}}\cdot{{\bf s}_{{\bf i},\alpha}}}, 
\label{hund}
\end{equation}
\vspace{-0.2cm}
\begin{equation}
H_{\rm Heis}= J_{{\rm NN}}\sum_{\langle{\bf ij}\rangle} {\bf S}_{{\bf i}}\cdot{\bf S}_{{\bf j}}
+J_{{\rm NNN}}\sum_{\langle\langle{\bf im}\rangle\rangle} {\bf S}_{{\bf i}}\cdot{\bf S}_{{\bf m}}, 
\label{heis}
\end{equation}
\vspace{-0.2cm}
\begin{equation}
H_{\rm SL}=-g\sum_{\bf i}\Psi_{\bf i}\epsilon_{\bf i}, 
\label{sl}
\end{equation}
\vspace{-0.2cm}
\begin{equation}
H_{\rm OL}=\lambda\sum_{\bf i}\Phi_{\bf i}\epsilon_{\bf i}, 
\label{ol}
\end{equation}
\vspace{-0.2cm}
\begin{equation}\begin{split}
H_{\rm Stiff}= {1\over{2}}k\sum_{\bf i}\sum_{\nu=1}^4(|{\bf R}^{\bf i \nu}_{Fe-As}|-R_0)^2+\\
+k'\sum_{<{\bf ij}>}[({a_0\over{R^{\bf ij}_{Fe-Fe}}})^{12}-
2({a_0\over{R^{\bf ij}_{Fe-Fe}}})^6].
\label{stiff}
\end{split}
\end{equation}
\vspace{-0.2cm}

\subsection{Lattice Distortions}

\begin{figure}[thbp]
\begin{center}
\subfigure {\includegraphics[trim = 10mm 5mm -2mm 5mm,width=0.3\textwidth,angle=0]{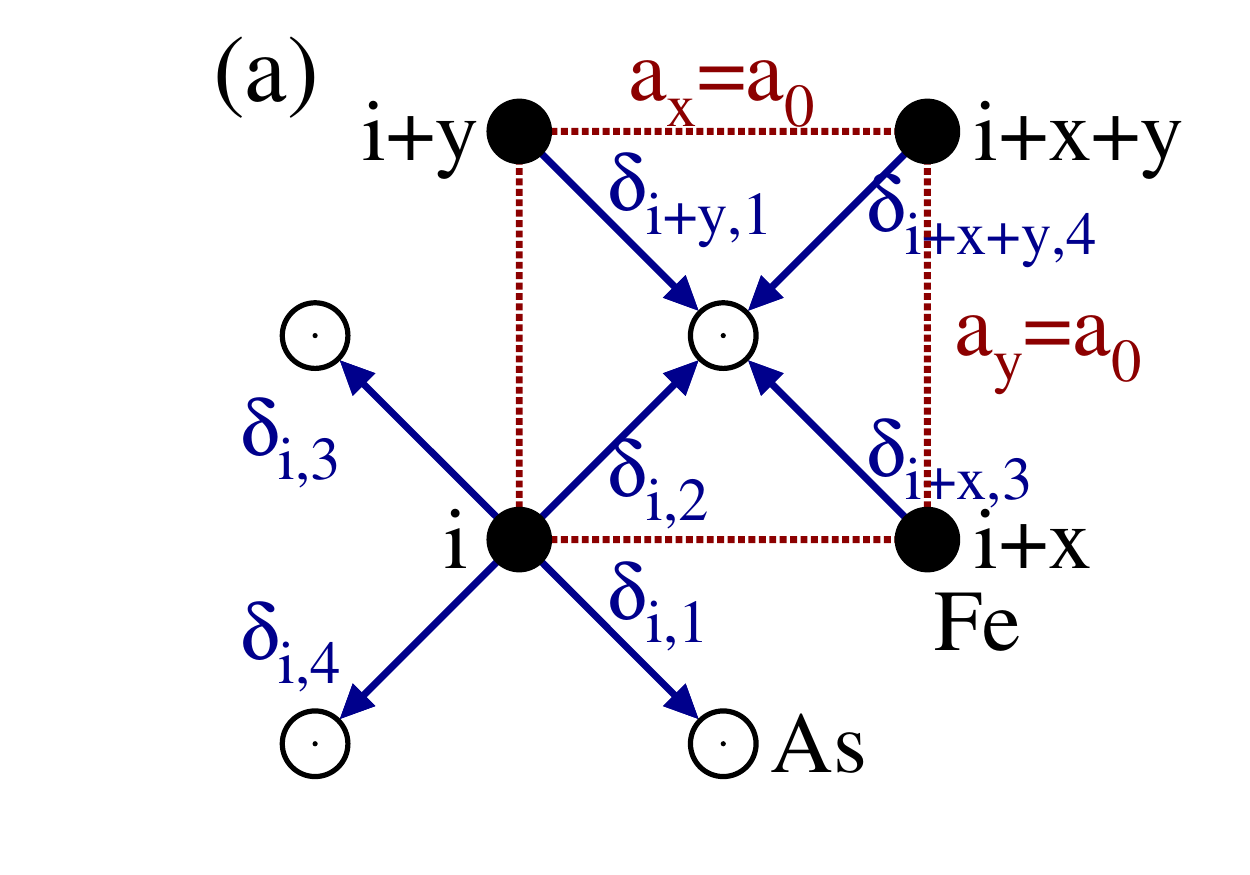} }
\subfigure {\includegraphics[trim = 6mm 3mm -1mm 7mm,width=0.15\textwidth,angle=0]{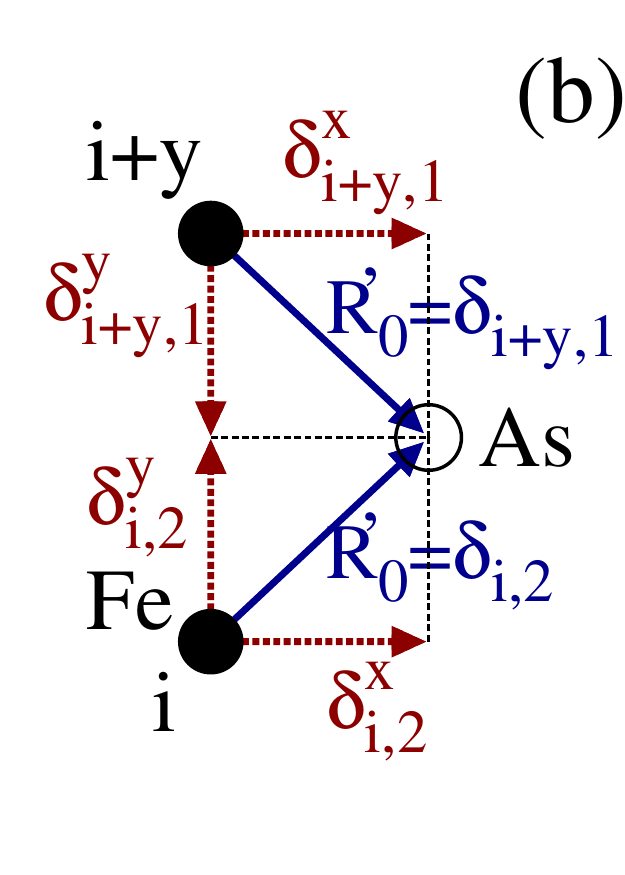} }\\
\vspace{-0.3cm}
\caption{(color online) (a) Schematic representation of the equilibrium 
positions of the Fe-As lattice (projected on the $x-y$ plane). Four Fe atoms 
are indicated with filled circles and labeled by their site index ${\bf i}$ 
(with ${\bf x}$ and ${\bf y}$ being unit vectors along the axes).
The open circles indicate the projection of the equilibrium position of the As 
ions in the $x-y$ plane. The distance between an Fe atom at site ${\bf i}$ 
and its four neighboring As atoms is indicated by $\delta_{{\bf i},\nu}$ with $\nu$ 
running from 1 to 4 as shown (blue arrows). 
In equilibrium, $\delta_{{\bf i},\nu}=R'_0=\sqrt{2}a_0/2$ where $R'_0$ is 
the projection on the $x-y$ plane of $R_0$, 
the equilibrated Fe-As distance. 
The red dashed lines indicate the case $a_x=a_y=a_0$, namely 
the equilibrium distance between neighboring Fe atoms. 
(b) Schematic representation of the variables $\delta_{{\bf i},\nu}^x$ 
and $\delta_{{\bf i},\nu}^y$ (red arrows) for the case (${\bf i},2$) and 
(${\bf i+y},1$) in the equilibrium configuration.}
\label{lattice}
\end{center}
\end{figure}

\begin{figure}[thbp]
\begin{center}
\subfigure {\includegraphics[trim = 10mm 5mm -2mm 5mm,width=0.3\textwidth,angle=0]{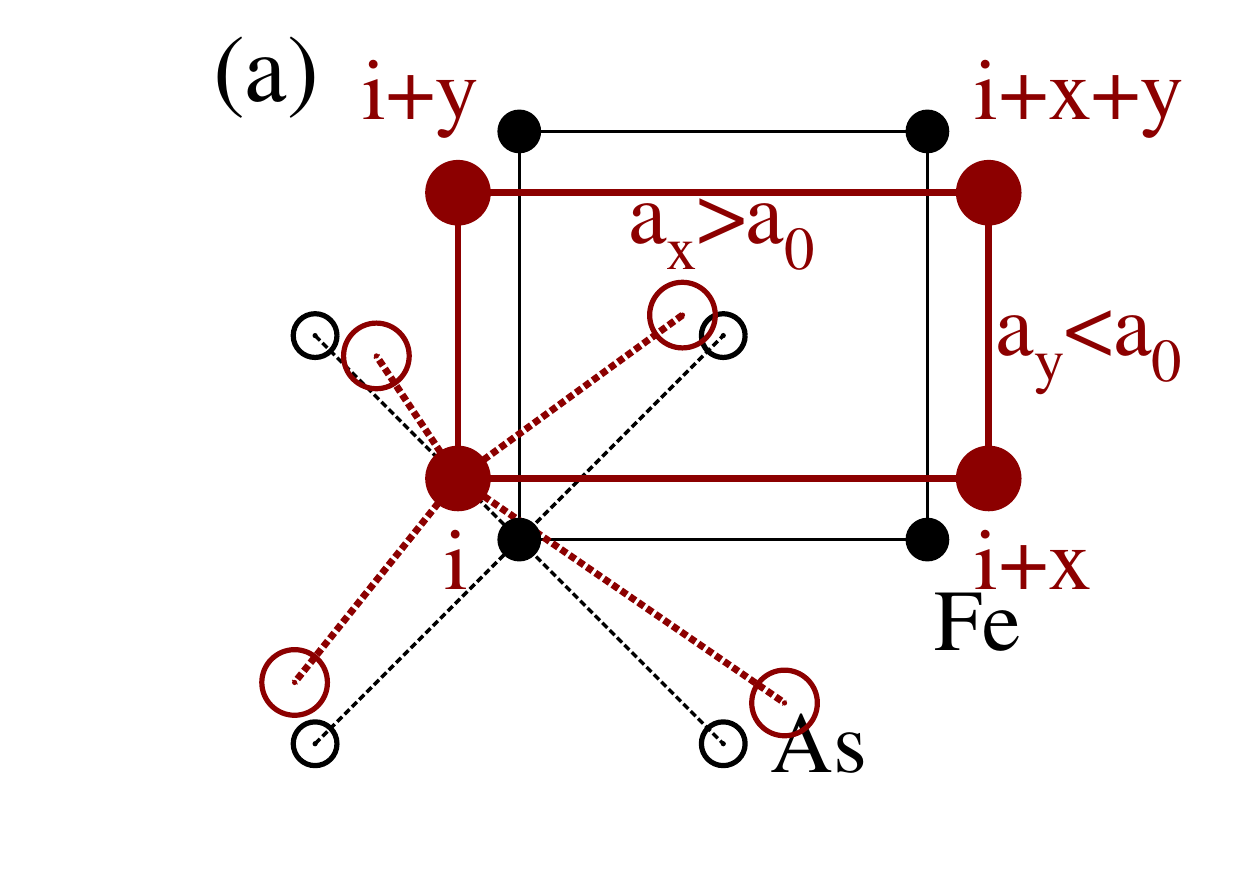} }
\subfigure {\includegraphics[trim = 6mm 3mm -1mm 7mm,width=0.15\textwidth,angle=0]{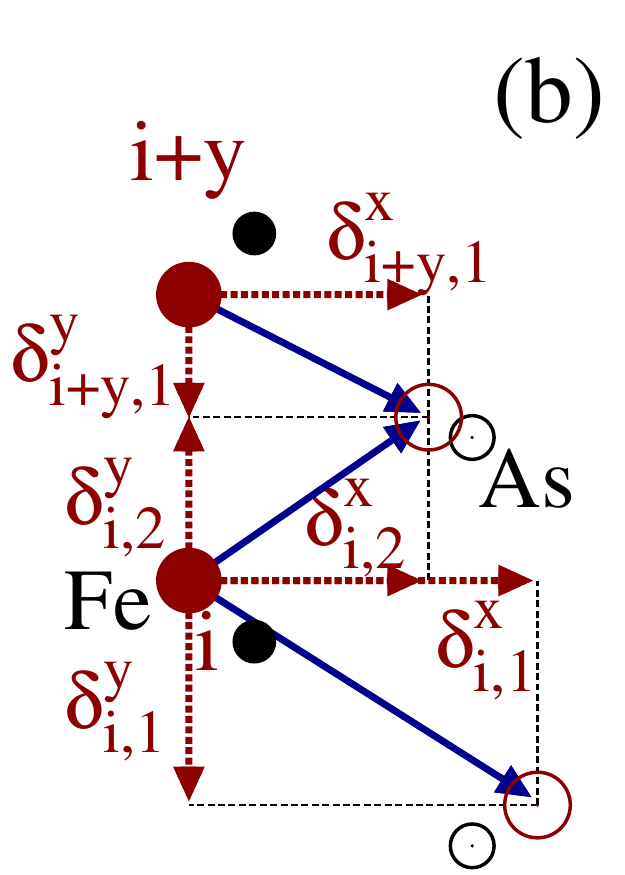} }\\
\vspace{-0.3cm}
\caption{(color online) (a) Schematic representation of a non-equilibrium 
position of the Fe-As  lattice (projected on the $x-y$ plane). 
Four Fe atoms are indicated with red filled circles and labeled 
by their site index ${\bf i}$ 
(with ${\bf x}$ and ${\bf y}$ being unit vectors along the axes).
The red open circles indicate the projection 
on the $x-y$ plane of the non-equilibrium position of the As atoms. 
The distance between neighboring Fe atoms is $a_x$ ($a_y$) along $x$ ($y$) 
indicated by red lines. 
The red dashed lines show $R'^{{\bf i}\nu}_{Fe-As}$, which is the 
projection on the plane $x-y$ of the Fe-As distance $R^{{\bf i}\nu}_{Fe-As}$.
The equilibrium position of the atoms 
is indicated by the black symbols. 
(b) Schematic representation of the variables $\delta_{{\bf i},\nu}^x$   and $\delta_{{\bf i},\nu}^y$  
(red arrows) for (${\bf i},1$), (${\bf i},2$), and 
(${\bf i+y},1$) in an out-of-equilibrium configuration.
The variables $\delta_{{\bf i},\nu}^x$ and $\delta_{{\bf i},\nu}^y$ 
are the $x$ and $y$ components of the distance between Fe and As atoms,
$R^{{\bf i}\nu}_{Fe-As}$, between the non-equilibrium position of the 
Fe atom at site ${\bf i}$ (filled red circle) and the As atom labeled by
(${\bf i},\nu$) (open red circle). The corresponding equilibrium 
positions are indicated by the black symbols.}
\label{outeq}
\end{center}
\end{figure}

The definition of the lattice variables used in our calculations
is shown in Fig.~\ref{lattice}. Panel (a) indicates the equilibrium 
position of the Fe atoms at the sites ${\bf i}$ of a square lattice, 
with equilibrium lattice constants $a_x=a_y=a_0$. In equilibrium, namely without
the influence of the electronic degrees of freedom,
the As atoms are separated by $a_0/2$ from each Fe
with regards to their $x$ and $y$ axes coordinates, while they are at distance $a_0/2$
above or below the $x-y$ plane on alternating plaquettes (remember the As atoms
are not in the same plane as the Fe atoms). 
During the Monte Carlo simulation the As atoms are allowed to move {\it locally} away 
from their equilibrium positions but with movements restricted to be 
only along the $x$ and $y$ directions for simplicity. The Fe atoms, on
the other hand, can only move {\it globally} also along the $x$ $(y)$ direction
[see panel (a) of Fig.~\ref{outeq}] such that the inter-Fe distance 
$a_x$ ($a_y$) arises from the constraint
$2Na_r=\sum_{{\bf i}=1}^N \sum_{\nu}|\delta_{{\bf i},\nu}^r|$. In this
formula $r=x,y$, $N$ is the number of sites of the lattice, and 
$\delta_{{\bf i},\nu}^r$ is the component along the $r$ axis of the distance
between the Fe atom at site ${\bf i}$ and one of the As atoms  
in the neighboring plaquette. The four As neighbors to a given Fe are 
labeled by the index $\nu=1,...,4$.
The equilibrium values of $\delta_{{\bf i},\nu}^r$ are shown 
in panel (b) of Fig.~\ref{lattice} while  
non-equilibrium values are shown in panel (b) of Fig.~\ref{outeq}.
In the latter, the atomic equilibrium positions
are shown in black and the non-equilibrium positions in red. 

\begin{figure}[thbp]
\begin{center}
\subfigure {\includegraphics[trim = 10mm 5mm 30mm 5mm,width=0.22\textwidth,angle=0]{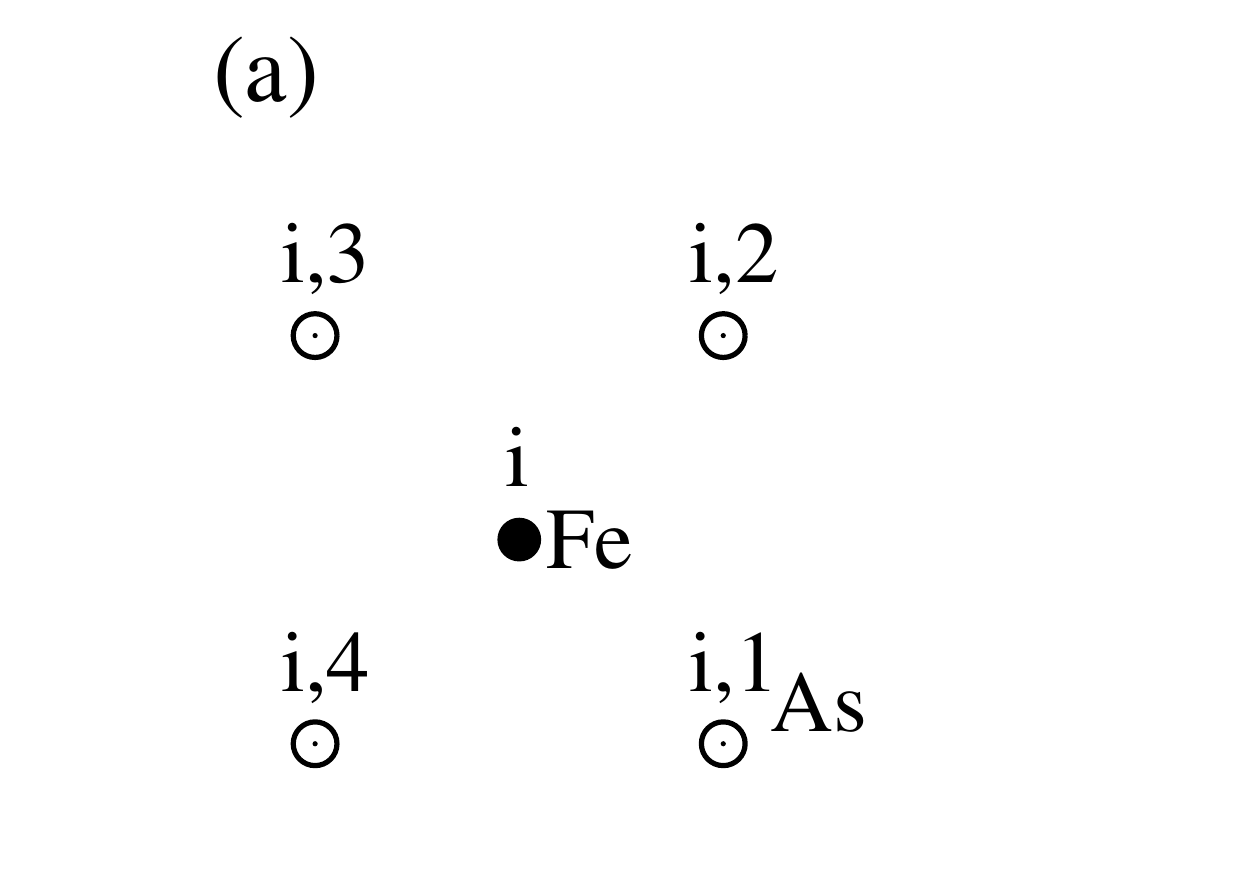} }
\subfigure {\includegraphics[trim = 20mm 5mm 20mm 5mm,width=0.22\textwidth,angle=0]{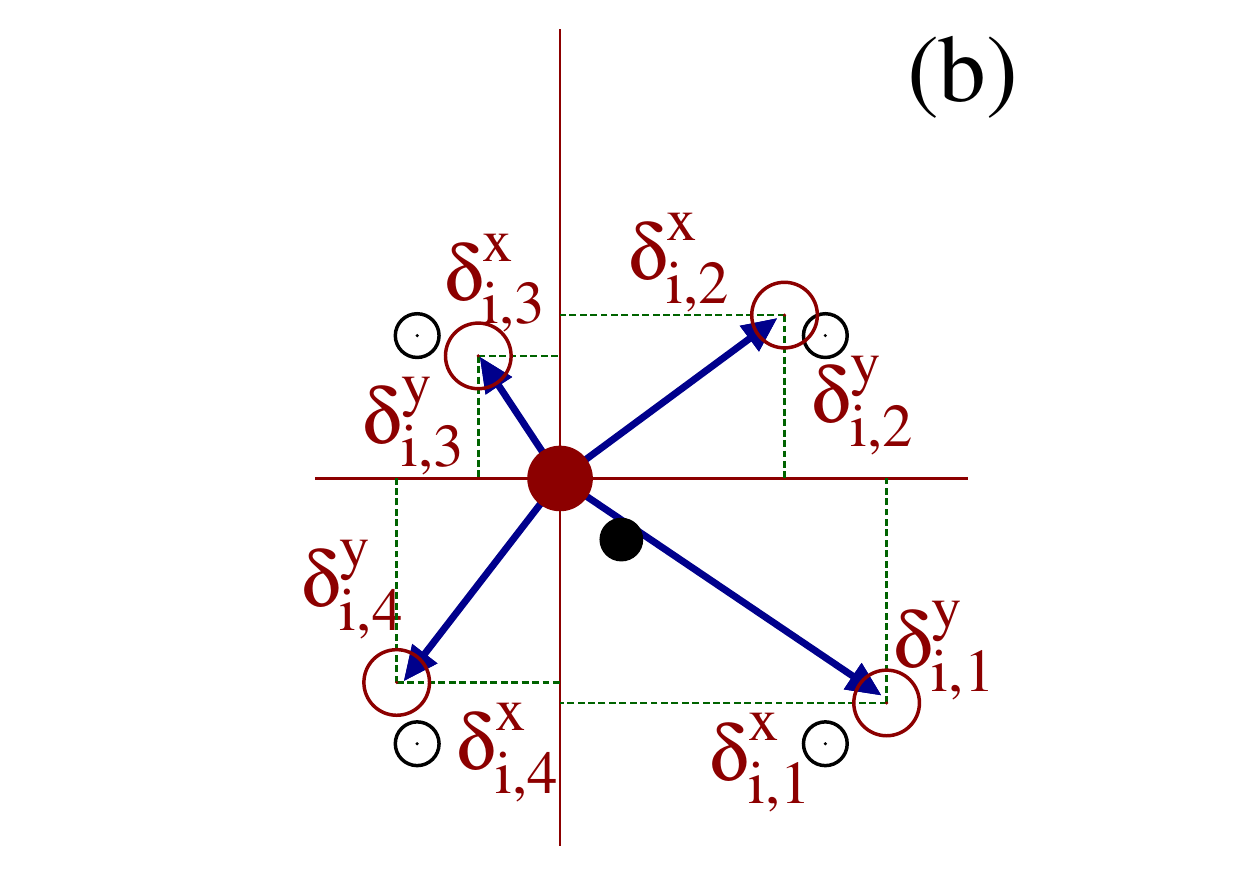} }\\
\vspace{-0.3cm}
\caption{(color online) (a) Schematic representation of the equilibrium 
position of the Fe atom at lattice site ${\bf i}$ (filled circle) surrounded 
by the four As atoms at their
equilibrium positions on the $x-y$ plane labeled by the index $\nu=1, 2, 3, 4$. 
(b) Schematic representation of the 
variables $\delta_{{\bf i},\nu}^x$ and $\delta_{{\bf i},\nu
}^y$ that define the $x$ and $y$ components of the distance 
between the non-equilibrium position of the Fe atom 
at site ${\bf i}$ (filled red circle) and its four
neighboring As atoms (open red circles). The corresponding equilibrium 
positions are indicated by the black symbols.}
\vspace{-0.4cm}
\label{displacement}
\end{center}
\end{figure}

The $\mathcal{O}_{rth}$ strain $\epsilon_{\bf i}$ defined in Eq.~\ref{epsilon} 
is schematically shown in panel (b) of Fig.~\ref{displacement} 
where the displacements $\delta_{{\bf i}\nu}^r$'s 
at site ${\bf i}$ are shown for $r=x,y$ and 
$\nu=1,2,3,4$, while panel (a) depicts 
the undistorted lattice as reference.

\subsection{Monte Carlo Technique}

The Monte Carlo technique used here to study the 
spin-fermion model defined in Eq.~1 is standard and it has been
extensively discussed in previous publications~\cite{CMR,shuhua} 
that can be consulted by the reader for more details.
In this technique, the acceptance-rejection MC steps 
are carried out visiting the classical spins one by one and the classical
lattice degrees of freedom also one by one. At each of these steps 
a full diagonalization of the fermionic hopping term $H_{\rm Hopp}$ in the background
of the classical spin and lattice d.o.f.
is carried out via library subroutines to calculate the energy that
enters in the Metropolis algorithm. This frequent diagonalization 
renders the technique rather time consuming. For this reason the
simulation is here limited to 8$\times$8 clusters. For the MC time
evolution, the previously described Hamiltonian $H_{\rm SF}$ is used  
with periodic boundary conditions.
However, for the measurement of observables
``twisted boundary conditions'' (TBC) are employed~\cite{salafranca}. In the case of 
TBC the classical
spin and lattice configurations are assumed replicated in space with
a difference of a phase factor such that a better resolution is achieved
with regards to the wavevector ${\bf k}$. The reason is that a larger lattice
(the replicated one) contains more eigenstates and gives a more continuous
distribution of eigenvalues, reducing size-effects. In practice, TBC are
introduced via phase factors $\phi$ that are added in the hopping amplitudes,
schematically denoted by $t$ (in reality, there are several different hopping
amplitudes connecting NN and NNN Fe sites and their several orbitals, but for all
of them the same phase factor must be used). The TBC amounts to replacing $t$ by
$e^{i\phi} t$, with $\phi = 2\pi m /M$ where $m = 0, 1, ..., M-1$ and the number
of possible wavevectors in the $x$ or $y$ directions becomes $L = 8 \times M$.

\subsection{Parameter values}

In this subsection, the actual values of the 
parameters used in the Hamiltonian Eq.~\ref{ham} are discussed. 
The dimensionless orbital-lattice (spin-lattice) coupling is given by 
$\tilde\lambda={\lambda\over{\sqrt{kt}}}$ ($\tilde g={g\over{\sqrt{kt}}}$), 
where $t$ is an effective hopping related to the bandwidth $W$ so 
that $t\approx W/4$. As a consequence, here it will be used
$\tilde\lambda={2\lambda\over{\sqrt{kW}}}$ ($\tilde g={2g\over{\sqrt{kW}}}$) 
with $W$=$3$~eV which is the bandwidth for the three-orbital model~\cite{three}. 
The estimation of $\tilde\lambda$ in previous
literature ranges from 0.1 to 0.8~\cite{kontani,boeri,ono}. Values in the range 
$\tilde\lambda=2\lambda/\sqrt{kW}=0-1.2$
have been used here, with the largest value only employed to highlight
the incorrect lattice distortion obtained in that limit. 
In our MC study a small value of $\tilde\lambda$ approximately 0.1 
is needed to observe a nonzero difference between $T_N$
and $T_S$ within our numerical resolution. 

The spin-lattice coupling has been estimated to be
$\tilde g=0.002-1$ in the literature~\cite{spin3,kontani2}.
In Ref.~\onlinecite{kontani2}, $g_{66}=\eta^2_{66}/C_{66,0}=g^2/k=0.12-0.21$ eV. 
Then $\tilde g=\sqrt{g_{66}/t}$. If $t\approx 0.2$~eV, 
then $\tilde g\approx 0.2-1$. 
But note that $g_{66}=3.4\times 10^{-6}$ eV
according to Ref.~\onlinecite{spin3} 
indicating $\tilde g\approx 0.004$, i.e., in previous efforts 
a wide range for the spin-lattice coupling has been discussed.

\begin{figure}[thbp]
\begin{center}
\subfigure {\includegraphics[trim = 8mm 5mm 4mm 5mm,width=0.22\textwidth,angle=0]{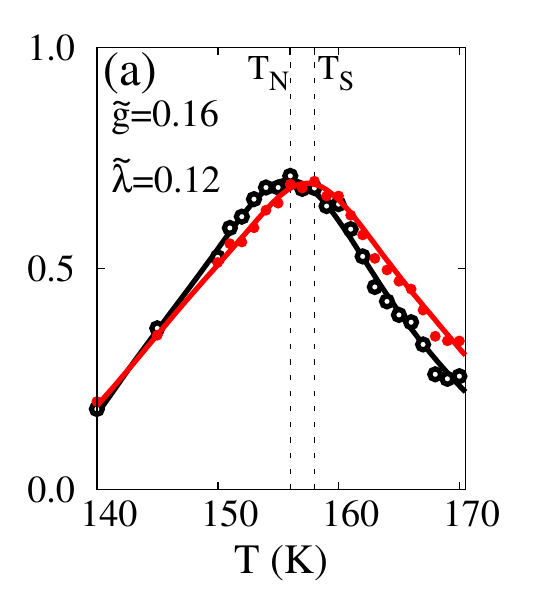}\label{g20l15}}
\subfigure {\includegraphics[trim = 4mm 5mm 8mm 5mm,width=0.22\textwidth,angle=0]{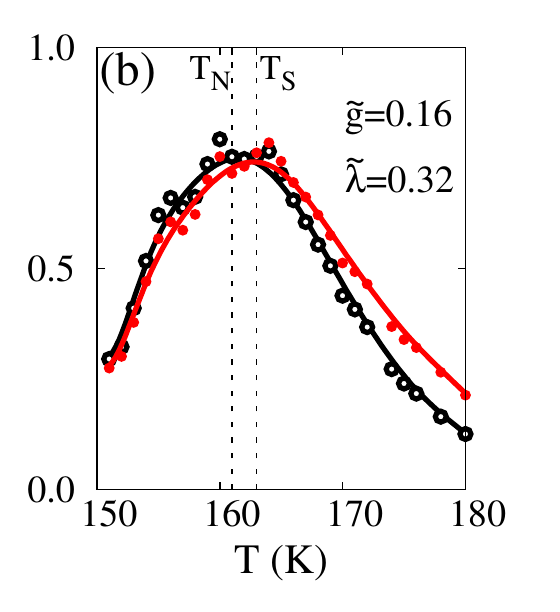}\label{g20l40}}\\
\subfigure {\includegraphics[trim = 8mm 5mm 4mm 5mm,width=0.22\textwidth,angle=0]{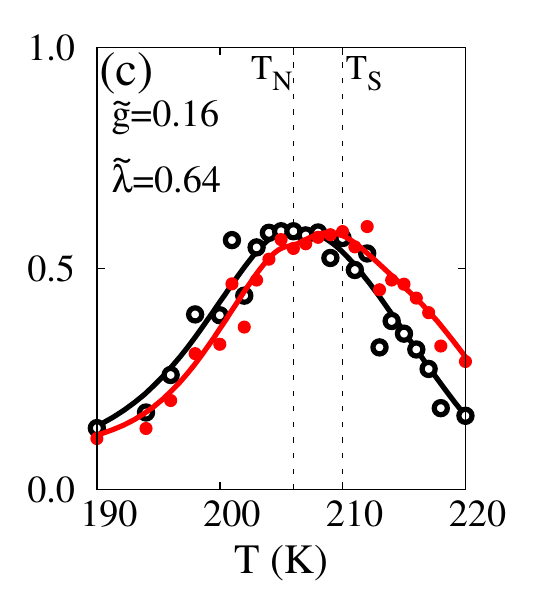}\label{g20l80}}
\subfigure {\includegraphics[trim = 4mm 5mm 8mm 5mm,width=0.22\textwidth,angle=0]{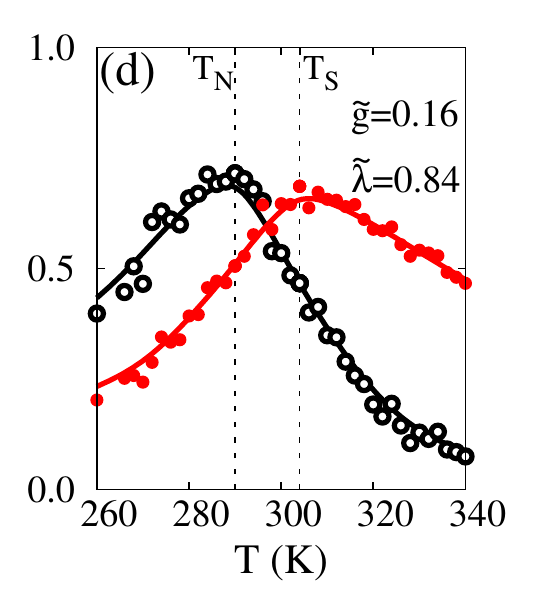}\label{g20l100}}\\
\vspace{-0.3cm}
\caption{(color online) $T_S$ and $T_N$ determined by 
susceptibility measurements during the Monte Carlo simulations. All results are 
at the couplings $\tilde g = 0.16$. The different panels correspond to 
(a) $\tilde \lambda = 0.12$,
(b) $\tilde \lambda = 0.32$,
(c) $\tilde \lambda = 0.64$, and
(d) $\tilde \lambda = 0.84$. The results in red (solid) points are for 
the lattice susceptibility $\chi_{\delta}$. The results in black (open) points
are for the spin susceptibility $\chi_{\rm S(\pi,0)}$.
}
\label{Fig10}
\end{center}
\end{figure}

\subsection{Determination of $T_N$ and $T_S$}

To determine the values of $T_S$ and $T_N$ extensive MC simulations 
and measurements of the magnetic and lattice 
susceptibilities, $\chi_S$ and $\chi_{\delta}$, were performed because of the
small difference between these critical temperatures at small values
of the couplings to the lattice. 
In the four panels of Fig.~\ref{Fig10} the susceptibilities 
that allowed us to determine the values for $T_S-T_N$ 
at the particular coupling $\tilde g=0.16$ are presented [see Fig.~\ref{g2l0}(b)]. 
Since at couplings such as $\tilde \lambda$=0.8 the difference between the
critical temperatures is clear, that give us confidence that in the region
of smaller $\tilde \lambda$'s the results are reliable since the difference
$T_S - T_N$ can be followed with continuity from large $\tilde \lambda$ 
to small $\tilde \lambda$.


\begin{thebibliography}{}
\bibitem{footep} According to the experimental results 
it is known that $\langle\epsilon_{\bf i}\rangle<0$ if the magnetic order of the ground state is $(\pi,0)$~\cite{fisher}. For this reason only positive values of 
the spin-lattice constant $g$ must be considered since 
otherwise an unphysical lattice distortion would result.
\bibitem{foot} If only harmonic terms are considered 
for the Fe atoms the results do not change but it takes 
longer to achieve numerical convergence.
\bibitem{f3} $W$=$3$~eV is the bandwidth of the three-orbital model~\cite{three}.
\bibitem{lambda} If in the orbital-lattice term $H_{\rm OL}$ 
a negative value of $\epsilon$ is introduced and kept fixed, 
so that $a_x>a_y$ as in experiments, 
a distortion of the FS that favors the intra-orbital nesting along 
$(0,\pi)$ is induced. Thus, changing the sign of the coupling in 
$H_{\rm OL}$ leads to the correct lattice distortion but the incorrect
spin order.
\end{thebibliography}
\end{document}